\documentclass[twocolumn]{aastex631}

\newcommand{\mytilde}{\raise.19ex\hbox{$\scriptstyle\sim$}}

\usepackage{floatrow}
\usepackage{amsmath}
\usepackage{subfigure}

\begin{document}


\title{Tracing the Formation History of Intrahalo Light with Horizon Run 5}

\correspondingauthor{M. James Jee}
\email{gudwls4478@yonsei.ac.kr, mkjee@yonsei.ac.kr}

\author[0000-0001-9139-5455]{Hyungjin Joo}
\affiliation{
Department of Astronomy,
Yonsei University, 50 Yonsei-ro, Seoul 03722, Republic of Korea}
 
\author[0000-0002-5751-3697]{M.James Jee}
\affiliation{
Department of Astronomy,
Yonsei University, 50 Yonsei-ro, Seoul 03722, Republic of Korea}
\affiliation{
Department of Physics,
University of California, Davis, One Shields Avenue, Davis, CA95616, USA}

\author[0000-0002-4391-2275]{Juhan Kim}
\affiliation{
Center for Advanced Computation,
Korea Institute for Advanced Study, 85 Hoegi-ro, Dongdaemun-gu, Seoul 02455, Republic of Korea}

\author[0000-0002-6810-1778]{Jaehyun Lee}
\affiliation{
Korea Astronomy and Space Science Institute (KASI), 
776 Daedeokdae-ro, Yuseong-gu, Daejeon 34055, Republic of Korea}

\author[0000-0002-9434-5936]{Jongwan Ko}
\affiliation{
Korea Astronomy and Space Science Institute (KASI), 
776 Daedeokdae-ro, Yuseong-gu, Daejeon 34055, Republic of Korea}
\affiliation{
University of Science and Technology, Gajeong-ro, Daejeon 34113, Republic of Korea}

\author[0000-0001-9521-6397]{Changbom Park}
\affiliation{
Center for Advanced Computation,
Korea Institute for Advanced Study, 85 Hoegi-ro, Dongdaemun-gu, Seoul 02455, Republic of Korea}

\author[0000-0001-5135-1693]{Jihye Shin}
\affiliation{
Korea Astronomy and Space Science Institute (KASI), 
776 Daedeokdae-ro, Yuseong-gu, Daejeon 34055, Republic of Korea}

\author[0000-0003-1414-1296]{Owain Snaith}
\affiliation{
GEPI, Observatoire de Paris, 
Université PSL, CNRS, 5 Place Jules
Janssen, 92190 Meudon, France}
\affiliation{
University of Exeter, 
School of Physics and Astronomy, Stocker Road, Exeter, EX4 4QL, UK}

\author[0000-0003-0695-6735]{Christophe Pichon}
\affiliation{
School of Physics, 
Korea Institute for Advanced Study, 85 Hoegiro, Dongdaemun-gu, Seoul 02455, Korea}
\affiliation{
CNRS and Sorbonne Université,
UMR 7095, Institut d’Astrophysique de Paris, 98 bis, Boulevard Arago, F-75014 Paris, France}
\affiliation{
IPhT,
DRF-INP, UMR 3680, CEA, L’Orme des Merisiers, B\^{a}t 774, F-91191 Gif-sur-Yvette, France}

\author[0000-0003-4446-3130]{Brad Gibson}
\affiliation{
University of Hull, Hull, Kingston upon Hull, GB}

\author[0000-0003-4164-5414]{Yonghwi Kim}
\affiliation{
Korea Institute of Science and Technology Information,
245 Daehak-ro, Yuseong-gu, Daejeon, 34141, Korea}

\begin{abstract}
We investigate the formation history of intrahalo light (IHL) 
using the high-resolution ($\mytilde1$~kpc), large-scale ($\mytilde1$~Gpc) cosmological hydrodynamical simulation, Horizon Run 5 (HR5).
IHL particles are identified by carefully considering both their binding energies and positions with respect to the tidal radii of individual galaxies.
By analyzing more than 1,200 galaxy groups and clusters with $\gtrsim10^{13}~M_{\sun}$
and tracing their individual IHL particles back in time,
we classify the origin of each IHL particle at each epoch based on the status of the originating galaxy into three categories:
brightest halo galaxy (BHG) formation/merger, satellite galaxy stripping, and pre-processing.
Our study reveals that the IHL production through BHG formation/merger is the predominant production channel, contributing over 60\% of the total IHL mass across all redshifts.
The second most significant IHL production channel is pre-processing, providing more than 20\% in the final HR5 snapshot. 
Stripping is negligible at $z>4$ but becomes gradually more important as halos mature at $z<4$.
Finally, we verify that IHL production through the disruption of dwarf galaxies and in-situ formation is negligible, contributing less than $\mytilde3$\% and $\mytilde$0.5\% to the total IHL production, respectively. 
\end{abstract}

\section{Introduction} \label{sec:intro}
Since \cite{Zwicky1937} first reported the detection of faint diffuse light between galaxies in the Coma cluster, numerous studies have explored its origin.
This faint light, originating from stars not gravitationally bound to individual galaxies within a cluster, has been referred to as intracluster light (ICL).
More recently, similar low-surface brightness features have also been observed on group scales, referred to as intragroup light \citep[IGL; e.g.,][]{Girardi2023, Martinez-Lombilla2023}. 
In this paper, we adopt the term intrahalo light (IHL) to encompass both ICL and IGL, providing a unified framework for analyzing these phenomena across different scales.

IHL has the potential to independently trace the global distribution of dark matter within halos, as suggested by both observational \citep[e.g.,][]{Jee2010, Montes2019, Kluge2024} and theoretical \citep[e.g.,][]{Asensio2020, Shin2022, Yoo2024} studies.
Furthermore, \cite{Deason2021} and \cite{Gonzalez2021} discussed theoretical and observational cases, respectively, where IHL may be utilized to measure the splashback radii of halos.

However, the exact justification for why IHL can serve as a luminous tracer of dark matter and the degree of the similarity remain unclear without a detailed understanding the formation channels of IHL.
If the majority of IHL stars form at high redshift, they have ample time to travel extensively within the halo and trace the global halo dark matter distribution.
On the other hand, if most IHL stars are produced only recently, their distribution may not be representative of the global dark matter distribution.

Various scenarios have been suggested regarding the IHL production mechanisms.
For example, IHL can be produced during the formation of brightest cluster galaxies (BCGs), as stars become unbound through major mergers or violent relaxation in the central region of the cluster \citep[e.g.,][]{Conroy2007, Murante2007, Ko2018}.
Alternatively, the IHL can originate from stars that are tidally stripped from satellite galaxies during their interactions with the cluster potential or other member galaxies \citep[e.g.,][]{Rudick2009, Contini2019}.
In addition, the pre-processing within galaxy group, prior to their accretion onto the main cluster, has been proposed as another significant formation channel for the IHL, though its detailed explanation remains uncertain. \citep[e.g.,][]{mihos2004interactions, Rudick2006, Contini2014}.
The disruption of dwarf galaxies \citep[e.g.,][]{Purcell2007}, and in-situ star formation \citep[e.g.,][]{Puchwein2010, Barfety2022, ahvazi2024} are also suggested as non-negligible channels.
A number of studies have demonstrated that among these processes, the IHL production via the disruption of dwarf galaxies and in-situ star formation may be insignificant \citep[e.g.,][]{Melnick2012,DeMaio2015, DeMaio2018}.

The primary mechanism contributing to IHL production is still under debate.
Theoretically, for instance, while \cite{Murante2007}, based on a cosmological hydrodynamical simulation, demonstrate that the formation of ICL is mainly associated with the build-up of the BCG, \cite{Contini2014}, based on a semi-analytic model, claim that the ICL production is predominantly contributed by satellite galaxy stripping. 
Observational studies are similarly inconclusive regarding the main IHL production channels. 
For example, \cite{DeMaio2015}, using their analysis of observations of four clusters, argue that the negative color gradients cannot be explained if BCG formation is the dominant IHL production channel, supporting the stripping channel as the main IHL formation mechanism.
However, recently, \cite{Joo2023} show that the mean IHL fraction in 10 high-redshift clusters at $1\lesssim z \lesssim 2$  observed with Hubble Space Telescope/Wide Field Camera IR is consistent with local measurements, which is interpreted as indicating that the dominant IHL production cannot be via satellite galaxy stripping.
With the advent of the JWST, significant progress is expected on the observational front in the near future, particularly in the $z>1$ regime.

There are four critical issues when the origin of IHL is studied with numerical simulations.
First, there is a difficulty in identification of IHL particles.
Similar to observations,  unambiguously separating star particles from those gravitationally bound to galaxies is non-trivial.
Second, caution is needed when interpreting the results from a limited sample size.
IHL properties vary widely from halo to halo, depending on each halo's detailed assembly history.
Third, it is computationally challenging to trace the origins of individual IHL particles reliably in a large cosmological simulation, where the number of particles exceeds tens of billions.
Fourth, the differentiation of various IHL formation mechanisms is not clearly defined. 
For example, merging and stripping within a cluster are not two distinctive processes but rather points on a continuum, whose classification depends on when we perform the classification.
In addition, pre-processing may refer to the same merging and stripping processes occurring on a group scale.

In this study, we study the origin of IHL using the high-resolution ($\mytilde1$~kpc), large-volume ($1049\times 119 \times 127$~Mpc$^3$) cosmological hydrodynamical simulation, Horizon Run 5 (HR5, \citealt{Lee2021}), carefully addressing the aforementioned outstanding issues in simulation-based IHL studies.
The large-volume allows us to find more than 1,200 halos with $M_{200}>10^{13}M_{\sun}$, enabling a statistically robust study.
Also, its high mass ($\mytilde10^6~M_{\sun}$) and spatial ($\mytilde1$~kpc) resolutions effectively resolve small galaxies down to $\mytilde10^9M_{\sun}$, thereby reducing the risk of mistakenly identifying star particles in unresolved halos as IHL particles.
The baryonic feedback of HR5 is carefully configured and tuned to closely emulate the cosmic star formation rate history, a crucial factor that may significantly influence the production rate of IHL.
When selecting IHL star particles, we rely on the Physically Self-Bound (PSB) galaxy finder \citep[\texttt{PGalF};][]{Kim2023}, which is based on the PSB halo finder \citep{kim2006}. 
The \texttt{PGalF} algorithm identifies the star particles of galaxies using binding energies and tidal radii, objectively differentiating bound and unbound particles with physical criteria.
Most importantly, we trace individual unbound particles back in time to their originating galaxies.
By determining the status of the originating galaxies based on objective and quantitative criteria at the epoch of classification, we ascertain the production mechanism of the corresponding unbound particles.

Our paper is organized as follows.
In \textsection\ref{sec:data}, we provide a brief description of HR5 and our scheme for tracing IHL origin.
\textsection\ref{sec:result} presents the result. We discuss the interpretation of the results in \textsection\ref{sec:discussion} before we conclude in \textsection\ref{sec:concl}. 
$M_{200}$ refers to the spherical mass of a halo at the radius, which encloses 200 times the critical density of the universe.

\begin{figure*}
    \centering
    \includegraphics[width=0.91\textwidth]{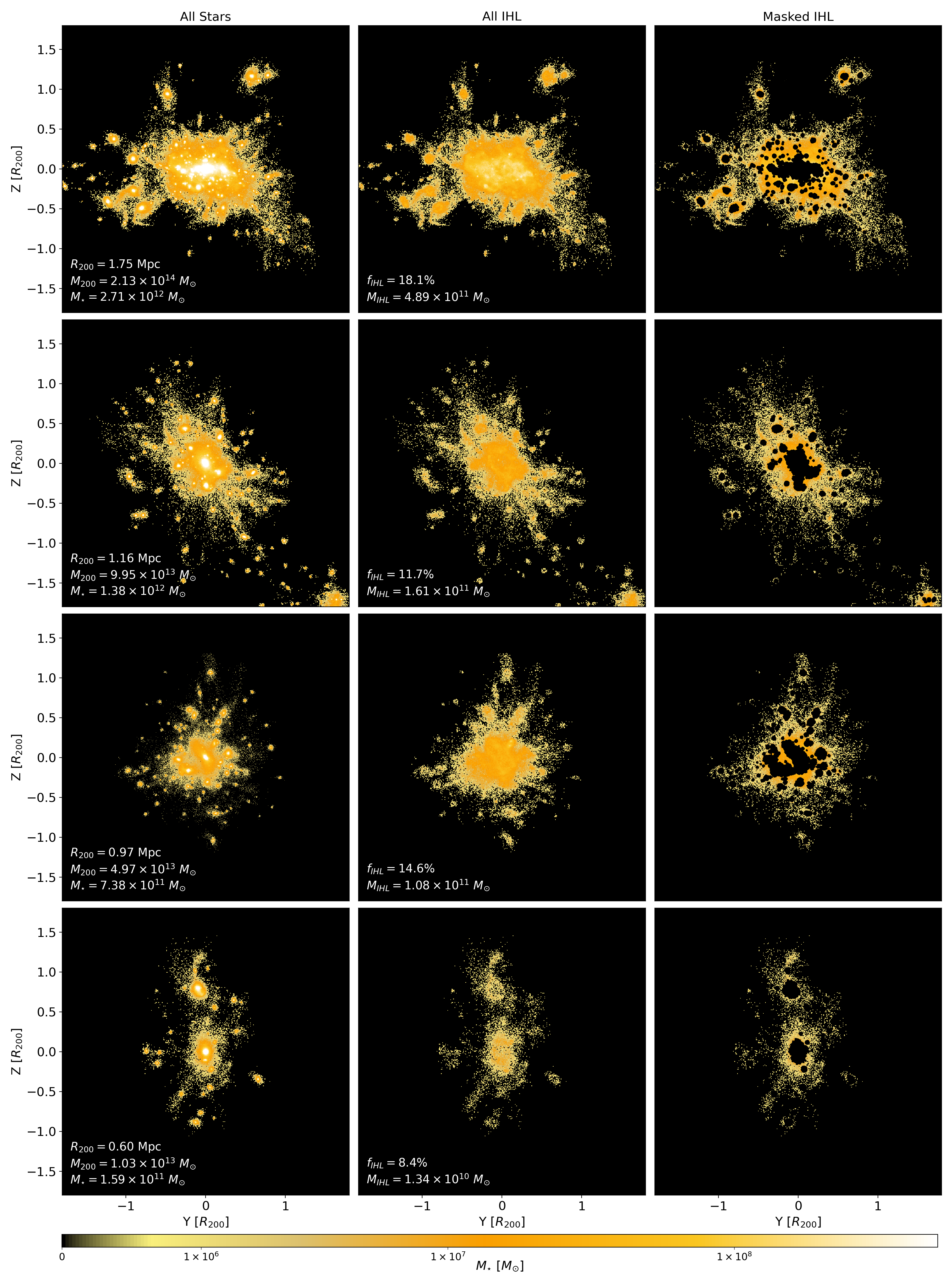}
    \caption{
    Stellar particle distributions in HR5 clusters. 
    Here, we select four exemplary halos at $z=0.625$ and display their stellar density distributions in the $y-z$ projection.
    Their $M_{200}$ values (from top to bottom) are $2.13 \times 10^{14}$, $9.95 \times 10^{13}$, $4.97\times 10^{13}$, and $1.03 \times 10^{13} ~ M_{\odot}$, respectively.
    Each row represents a different halo.
    The left column shows the density of all (bound+unbound) stellar particles, whereas the middle column displays only the density of unbound stellar particles.
    The images in the right column are produced by using bound particles as masks to depict observations, where masks are placed over galaxies.
    They show a number of low-surface brightness  features, including concentration near the BHGs, ongoing stripping, diffuse distribution associated with infalling groups, etc. 
    The red dashed lines indicate $r_{200}$.
    }
    \label{fig:halo_example}
\end{figure*}

\section{Simulation Data} \label{sec:data}
\subsection{Horizon Run 5} \label{subsec:HR5}
HR5 is a high-resolution, large-volume cosmological hydrodynamical simulation conducted using a modified version of the adaptive mesh refinement code RAMSES \citep{Teyssier2002}.
Within the parent $\mbox{(1.049 Gpc)}^3$ cubic volume, a cuboid of $1049 \times 119 \times 127$~Mpc$^3$ is set up to attain a maximum resolution of about 1 kpc (proper length) while simulating the large-scale structure up to $\mytilde1$~Gpc.
In HR5, \cite{Lee2021} implemented OpenMP into the  RAMSES code, which originally only utilized MPI parallelism.
The cosmological parameters adopted in HR5 are as follows:
$\Omega_{m}=0.3,\ \Omega_{b}=0.047,\ \Omega_{\lambda}=0.7,\ H_{0} = 68.4\ \mbox{km/s/Mpc}$, and $\sigma_{8}=0.816$ \citep{planck2016}.
The simulation stops at $z=0.625$, providing more than 1,200 halos with $M_{200}\gtrsim 10^{13}M_{\sun}$.
We traced the IHL particles in these halos and their progenitors back in time and identified their origins at $z \ge 0.625$.
Figure \ref{fig:halo_example} shows the stellar components of four exemplary halos at $z = 0.625$.
Their $M_{200}$ values from top to bottom are $2.13 \times 10^{14}$, $9.95 \times 10^{13}$, $4.97\times 10^{13}$, and $1.03 \times 10^{13} ~ M_{\odot}$, respectively.
The method for identifying IHL stars is described in \textsection\ref{subsec:halofind}. 

The HR5 subgrid physics is tuned to reproduce the cosmic star formation history constrained by observations.
The HR5 simulation incorporates a comprehensive set of subgrid physical models to capture key processes relevant to galaxy formation.
These include radiative gas cooling and heating with UV background radiation and reionization, star formation following a stochastic model, and feedback from supernovae and active galactic nuclei (AGN).
Chemical enrichment from Type II and Ia supernovae is also modeled, tracking the abundances of oxygen and iron.
Black hole growth is implemented via gas accretion, and AGN feedback is applied in a dual-mode scheme, combining thermal and kinetic modes. These physical models  ensure that HR5 reproduces large-scale structure while resolving the baryonic processes essential for galaxy evolution.
These modeling choices can influence the predicted properties of IHL. 
For instance, AGN feedback suppresses star formation in massive halos and can affect the spatial distribution and quantity of unbound stars contributing to the IHL. 
Similarly, the implementation of supernova feedback and chemical enrichment impacts the formation history and stellar population properties of galaxies, which in turn shape the structure and composition of IHL.
While our focus is on tracing the formation history of IHL, it is important to recognize that the detailed physical modeling adopted in HR5 may result in differences from other simulations with varying subgrid prescriptions. 
We therefore interpret our results within the context of the HR5 model framework.
One can find further details on HR5 in \cite{Lee2021}.

\subsection{Identification of Halos, Galaxies, and IHL} \label{subsec:halofind}
Halos are identified through an extended Friend-of-Friend (FoF) method using the following linking length:
\begin{equation}
l_{link}=0.2 \left ( \frac{m_p}{\Omega_m \rho_c} \right )^{1/3},    
\end{equation}
where $\rho_c$ and $m_p$ are the critical density at $z = 0$ and the particle mass, respectively.
This FoF scheme is applied to the mixture of different particles (dark matter, star, gas, and black hole). Two particles of different masses $m_1$ and $m_2$ are linked when they are closer than the average linking length: $l_{ave}=(l_1+l_2)/2$.
This results in a final sample of 1,202 halos 
with  $M_{200} > 10^{13} M_{\odot}$ in the final HR5 snapshot ($z=0.625$).
We adopt a mass threshold of $M_{200} > 10^{13}M_\odot$ at $z = 0.625$ in order to examine the formation of IHL in both group- and cluster-scale halos while maintaining a consistent evolutionary history.

Galaxies within an FoF halo are identified with \texttt{PGalF} \citep{kim2006,Kim2023}.
The \texttt{PGalF} algorithm carefully selects particles bound to local stellar or dark matter density peaks based on the total energy arguments and tidal radius criteria, ensuring that only physically self-bound structures are classified as galaxies.
Consequently, the remaining gravitationally unbound stars are classified as IHL.
Readers are referred to \cite{kim2006}, \cite{Lee2021}, and \cite{Kim2023} for details. 
Here we provide only a brief summary.

\texttt{PGalF} starts by identifying local density peaks defined by star particles.
To minimize the identification of spurious local peaks, two precautions are taken. One is
to merge multiple peaks if their distances are less than a threshold.
The other is to impose a minimum number of 25 stars and a minimum stellar mass of $10^7\ \mathrm{M_{\odot}}$ in the ``core region".
The core region is defined as the maximum volume enclosed by the isodensity surface that does not include another density peak.
The particles within each core region are considered galaxy candidate particles.
After the extraction of core particles surrounding the density peak, density cuts are applied to classify the remaining (non-core) particles based on the watershed algorithm.
The i$^{th}$ shell consists of particles between the i$^{th}$ and $(i+1)^{th}$ isodensity surfaces.
Finally, we examine the tidal boundary of each galaxy candidate and the total energy of particles within the boundary.
A particle bound to a galaxy cannot be its member if it is outside its tidal radius.

The middle column of Figure \ref{fig:halo_example} shows the IHL components identified with \texttt{PGalF} for the corresponding halos displayed in the left column. 
The images in the right column are obtained by masking out the regions occupied by bound particles, depicting the observational images where masks are placed over galaxies.
Figure \ref{fig:halo_example} reveals a number of low-surface brightness IHL features, including concentration near the BHGs, ongoing stripping, diffuse distribution associated with infalling groups, etc.

\begin{figure}
    \centering
    \includegraphics[width=\columnwidth]{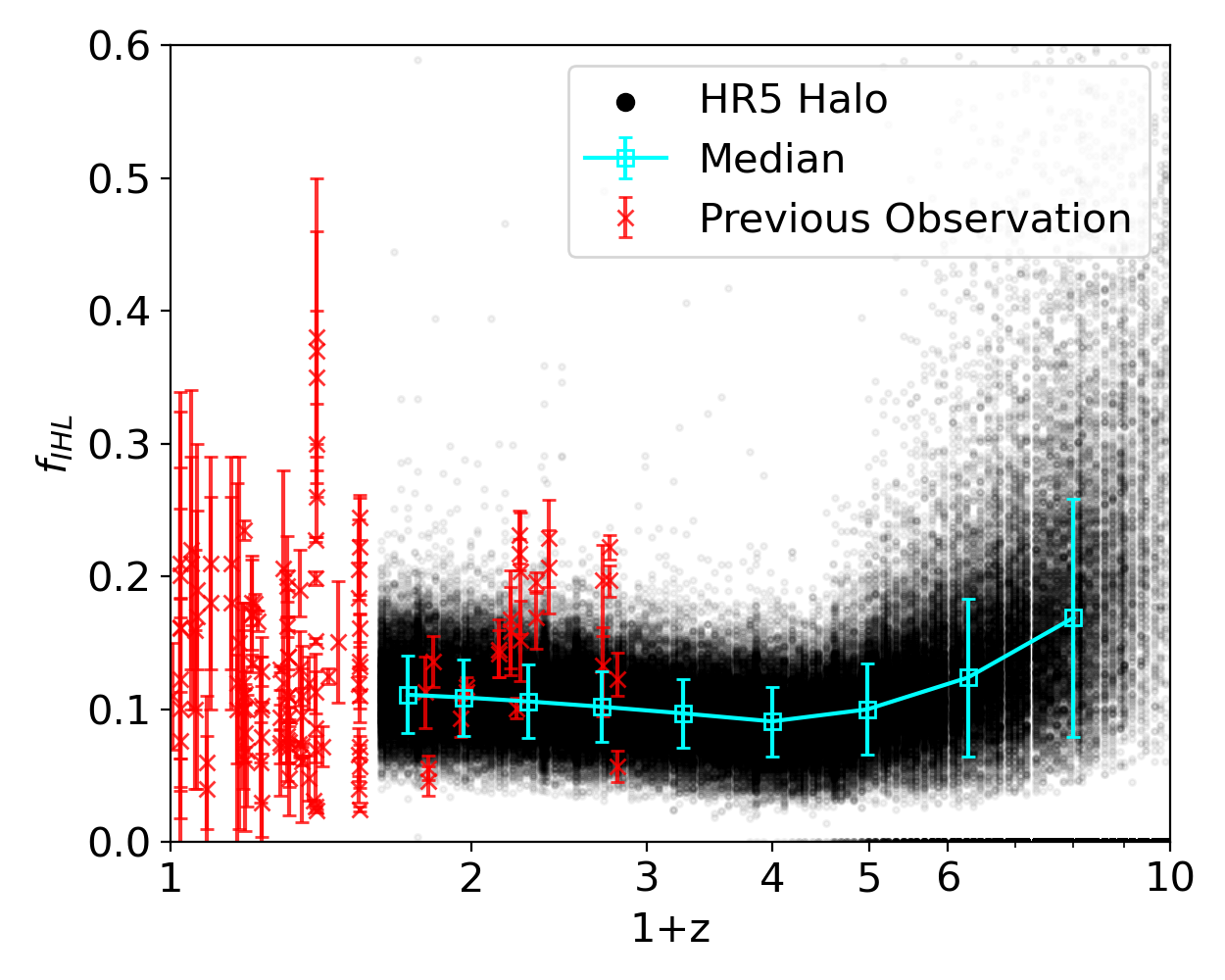}
    \vspace{0.5em}
    \includegraphics[width=\columnwidth]{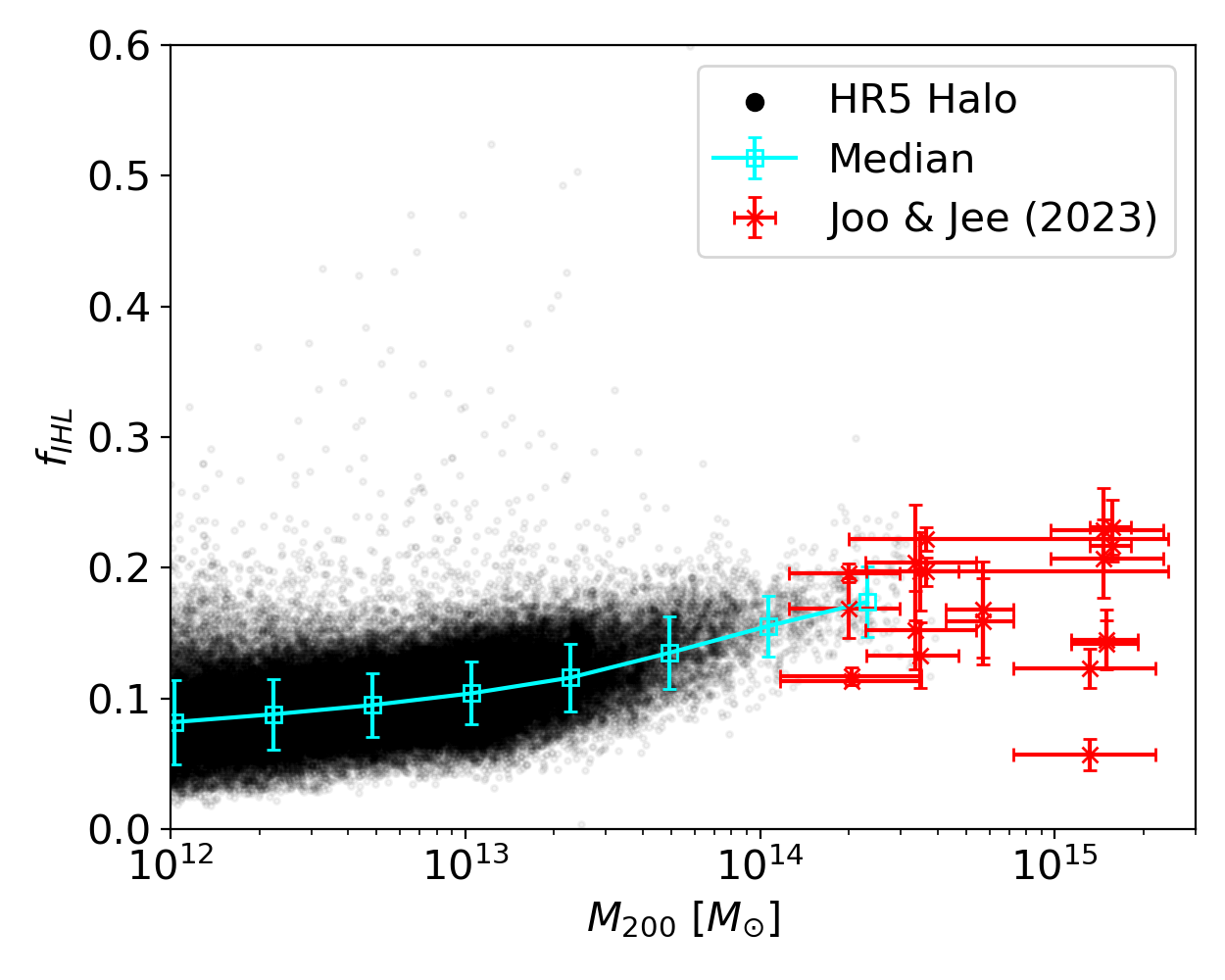}
    \caption{IHL fraction evolution from HR5.
    Black dots represent the values for individual halos.
    Cyan squares depict the medians, while the error bars indicate the standard deviations obtained within individual bins.
    The red crosses represent measurements from previous studies (see text for references).
    (Top) IHL fraction with evolution redshift.
    The IHL fraction shows moderate evolution in the $z\lesssim4$ regime, eventually reaching $f_{\mathrm{IHL}} \sim 0.12$.
    The fraction appears high at $z \gtrsim 4$, but this is uncertain due to unstable identification in poorly resolved halos.
    (Bottom) IHL fraction versus halo mass.
    $f_{\mathrm{IHL}}$ gradually increases with $M_{200}$,
    rising from $\sim 9 \%$ at $10^{12} M_{\odot}$ to $\sim 16 \%$ at $10^{14} M_{\odot}$.
    The HR5 prediction is consistent with the observations in \cite{Joo2023}.
    }
    \label{fig:fraction_IHL_comparison}
\end{figure}

\begin{figure*}
    \centering
    \includegraphics[width=\textwidth]{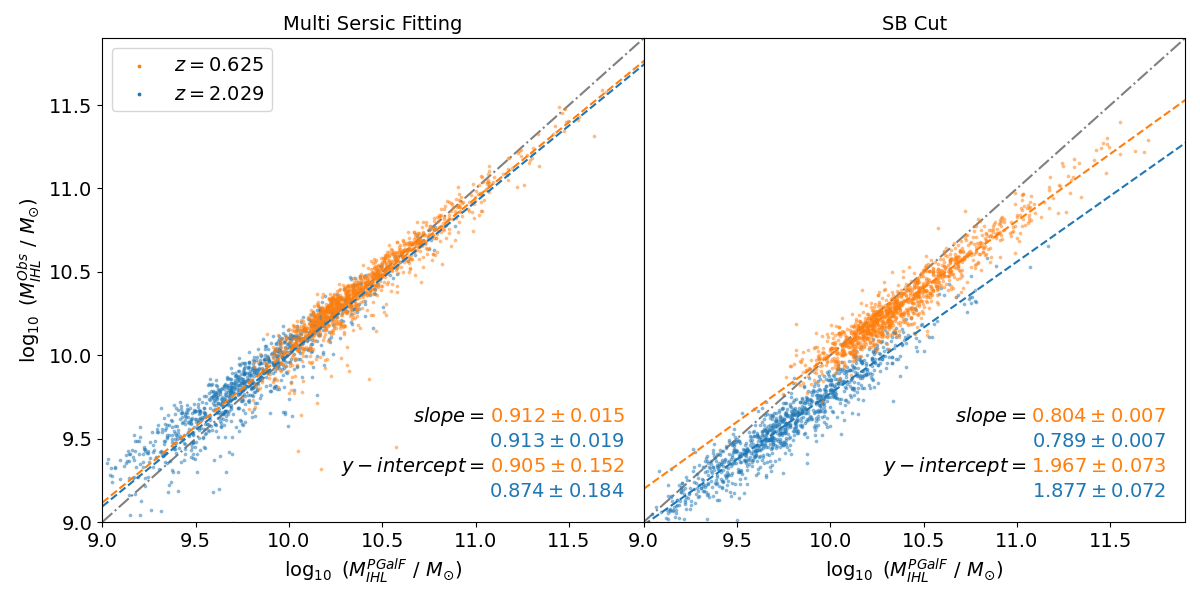}
    \caption{IHL recovery test for the multi-Sersic fitting (left) and surface brightness cut (right) methods.
    For the multi-Sersic fitting, we adopted the definition in \cite{Joo2023}.
    For the surface brightness cut method, we excluded the area where the surface brightness in the r-band is brighter than $26.5~\mbox{mag/arcsec}^2$.
    The $x$-axis represents the mass of IHL identified with \texttt{PGalF}, considered as ground truth, while the $y$-axis shows the recovered IHL mass.
    The gray dot-dashed line indicates one-to-one correlation.
    The multi-component fitting method is superior to the surface-brightness cut method, which exhibits a systematic offset between the low ($z=0.625$) and high ($z=2.029$) redshift results. 
    Also, a stronger mass dependency (i.e., the slope is significantly less than one) is observed when employing the surface brightness cut method.
    }
    \label{fig:method_IHL_comparison}
\end{figure*}

\subsection{Evolution of IHL Fraction} \label{subsec:fraction_evolution}

Before we begin extensive analysis on the origin of IHL based on IHL particle tracing, one useful quantity immediately available from the IHL identification is the IHL fraction as a function of redshift.
We define the IHL fraction as:
\begin{equation}
    f_{\mathrm{IHL}} = M_{\mathrm{IHL}}~/~M_{\mathrm{star}}
    \label{equation:ihl_frac}
\end{equation} 
where $M_{star}$ is the total stellar mass within the halo, and $M_{\mathrm{IHL}}$ is the total mass of IHL.
Since the flux ratio is commonly used to estimate the IHL fraction in observational studies, we first check the consistency between the mass and flux fractions before adopting the mass-based definition.
To estimate the flux from the HR5 data, we created mock observational images using the SKIRT radiative transfer code \citep{Camps2020}. 
We found that both the mass and flux ratios are highly consistent.
A comparison between the two prescriptions is presented in Appendix~\ref{appendix:ihl_fraction_comparison}.
Therefore, we adopt the mass ratio throughout.

Some theories predict that the IHL fraction rapidly decreases with increasing redshift, to the level of a few per cent at $z>1$ \citep[e.g.,][]{Rudick2011,Contini2014} while there are reports of conflicting results in observations \citep[e.g.,][]{Ko2018, Joo2023, Coogan2023, Werner2023}.
Since the IHL evolution is one of the critical diagnostics for constraining its origin, it is important to revise our understanding whenever new state-of-the-art numerical simulations become available. 

We present the IHL fraction evolution resulting from HR5 in Figure \ref{fig:fraction_IHL_comparison} and compare it with previous observational measurements including both mass fraction and flux fraction \citep{Ellien2019, Jimenez-Teja2018, Jimenez-Teja2019, Griffiths2018, Montes2018, Mihos2017, Alamo-Martinez2017, Morishita2017, Burke2012, Burke2015, Presotto2014, Jee2010, Krick2007, Feldmeier2004, Yoo2021, Joo2023}.
The halo-to-halo variation from our 1,202 halo sample is large, ranging from $\mytilde0.05$ to $\mytilde0.5$. 
The IHL fraction shows a mild evolution for decreasing redshift at $z\lesssim3$, with the median value reaching to $\sim 12\%$.
Observational studies show a slightly higher median of $\sim 17\%$ at $z \lesssim2$.

We further investigate the dependence of $f_{\mathrm{IHL}}$ on halo mass. 
The lower panel of Figure \ref{fig:fraction_IHL_comparison} shows that the $f_{\mathrm{IHL}}$ increases with halo mass.
Considering the mass dependence, the $\sim5\%$p difference in the IHL fraction between HR5 and observations can be explained by accounting for the difference in halo mass.
We defer the interpretation to \textsection\ref{sec:result}, where we discuss the origin of IHL based on particle tracing analysis.

\subsection{Surface Brightness Cut vs. Decomposition} \label{subsec:obs_method_comp}
Alongside the IHL fraction measurement, another useful quantity immediately available from the \texttt{PGalF} identification of IHL particles is the IHL measurement bias affecting observational studies. 
The traditional approach is the so-called surface brightness cut, where one considers diffuse light fainter than a threshold belonging to IHL. 
A more recent method is a multi-component decomposition technique \cite[e.g.,][]{ Janowiecki2010,Joo2023}, in which the diffuse light around the BHG is modeled using a superposition of multiple profiles (e.g., Sersic profiles).
The primary weakness of the first method is the arbitrariness of the threshold, which must be adjusted for different environments and distances.
The second approach is designed to overcome the weakness of the first method.
However, no extensive study has demonstrated that the diffuse light around the BHG can indeed be reliably decomposed with multiple profiles.

Figure \ref{fig:method_IHL_comparison} compares the true (output from \texttt{PGalF}) projected IHL mass with the quantities derived from the aforementioned two approaches.
For the surface bright cut method, we adopted
$m_{cut}=26.5~\mbox{mag / arcsec}^2$ in $r$, based on mock observations generated using the SKIRT.
The multi-component decomposition method provides consistent results between the low ($z=0.625$) and high ($z=2.029$) redshifts.
However, the surface brightness method shows a clear offset between the two redshifts.
In addition, we find that the slope in the multi-component fitting result is significantly closer to unity.
These differences are expected because a fixed value for the surface brightness cut neglects the dependence on the cluster mass and distance.
Therefore, we conclude that the multi-component fitting method is superior to the surface-brightness cut method when IHL is quantified from observations.

\subsection{Construction of Halo and Galaxy Chains}
\subsubsection{Halo Chain Construction}
Following halo and galaxy identification in individual snapshots, we locate the most massive galaxy (i.e., BHG) within each halo at $z= 0.625$.
Using its position and associated particle IDs, we find the parent halo in the previous snapshot.
We repeat this procedure until no parent halo is found in the higher-redshift snapshot, constructing the ``halo chain".

\begin{figure}
    \centering
    \includegraphics[width=\columnwidth]{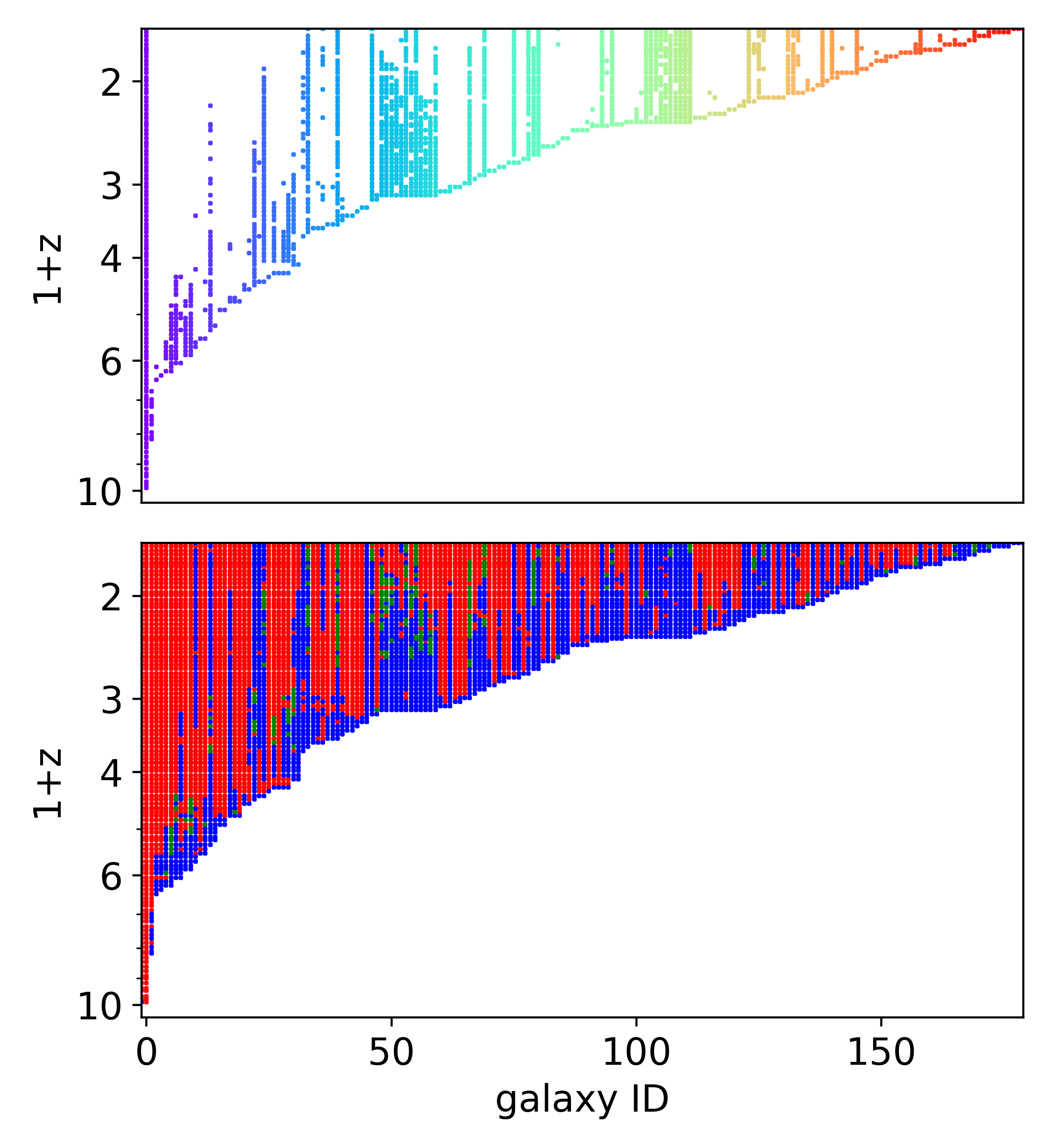}
    \caption{
    Galaxy chain construction and status classification. 
    Here, we chose a representative low-mass halo ($M_{200} = 2.41 \times 10^{13}M_\odot$) at $z=0.625$ to avoid visual clutter caused by large number of galaxies and stellar particles.
    We collected individual \texttt{PGalF} galaxy catalogs across snapshots and constructed galaxy chains based on stellar particle IDs.
    In the \texttt{PGalF} catalogs, galaxy IDs differ across snapshots, whereas in the galaxy chain, galaxy IDs are unique across snapshots.
    The lower the ID, the earlier it is chained (e.g., galaxy 1 appeared first and was followed by galaxy 2).
    (Top) We display the galaxy chains. 
    Different colors are used to depict different galaxies.
    A galaxy can disappear in the next snapshot as it may merge with another galaxy, belong to a different FoF halo, or fail to be resolved by the \texttt{PGalF} algorithm as an independent galaxy.
    For the same reason, it can reappear in subsequent snapshots.
    (Bottom)
    Classification of the galaxy status at each snapshot.
    The galaxies are classified as \texttt{M} (merged with the BHG) if indicated in red, \texttt{S} (stripped to the BHG) in green, and \texttt{P} (pre-processed) in blue.
    Sometimes, the status alternates between the classes because of the imperfect PGalF galaxy identification. 
    }
    \label{fig:galaxy_tree}
\end{figure}

\begin{figure*}
    \centering
    \includegraphics[width=0.88\textwidth]{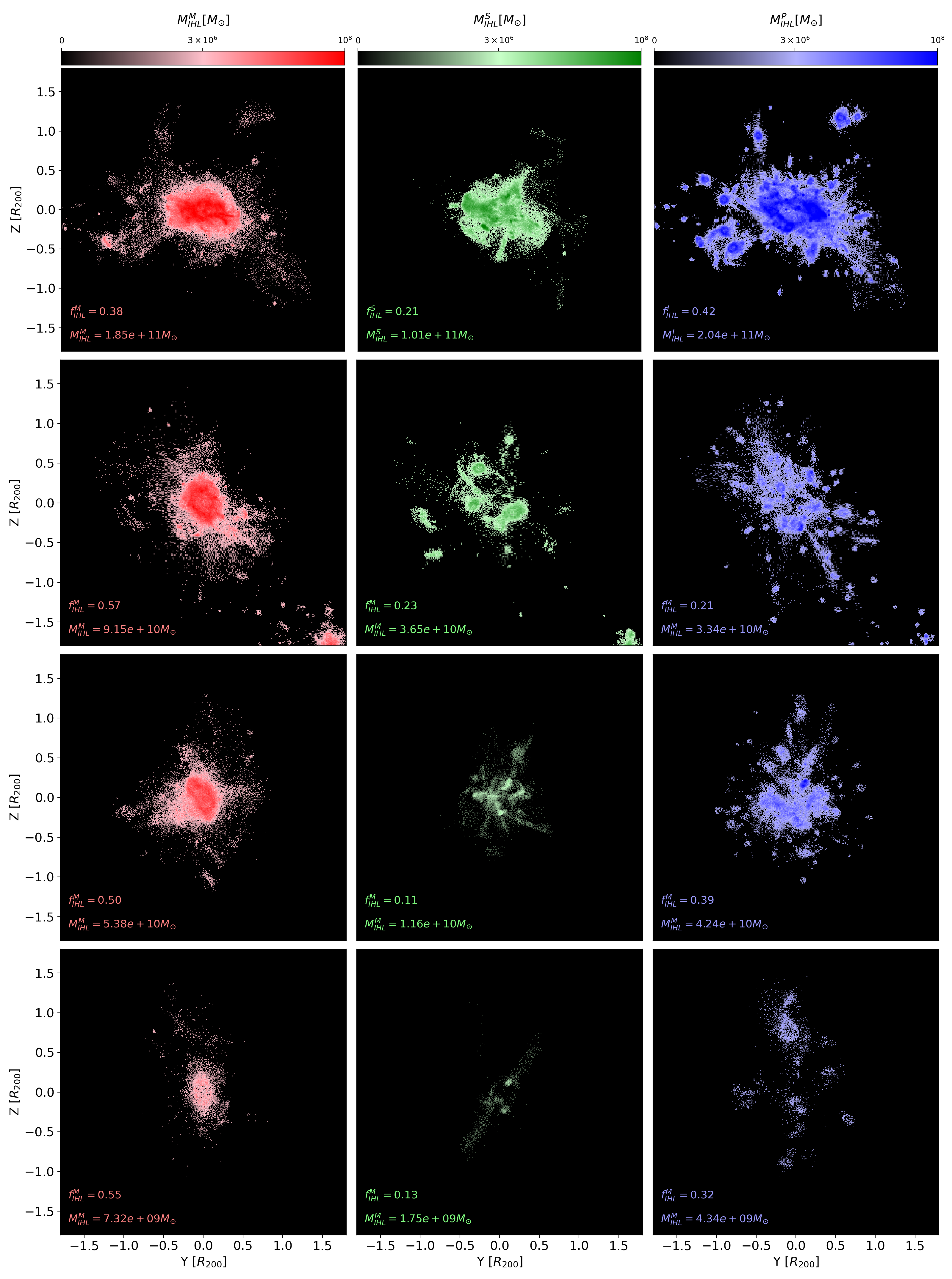}
    \caption{Spatial distribution of IHL particles with different origins.
    Each row displays the same halos illustrated in Figure \ref{fig:halo_example}.
    The left, middle, and right columns display the IHL particles classified as \texttt{M}, \texttt{S}, and \texttt{P}, respectively. 
    The \texttt{M} particles are concentrated near the BHGs.
    The distribution of \texttt{S} exhibits stripped features around the BHGs, whereas the \texttt{P} particles are found around the infalling galaxies.}
    \label{fig:class_example}
\end{figure*}

\begin{figure}
    \centering
    \includegraphics[width=\columnwidth]{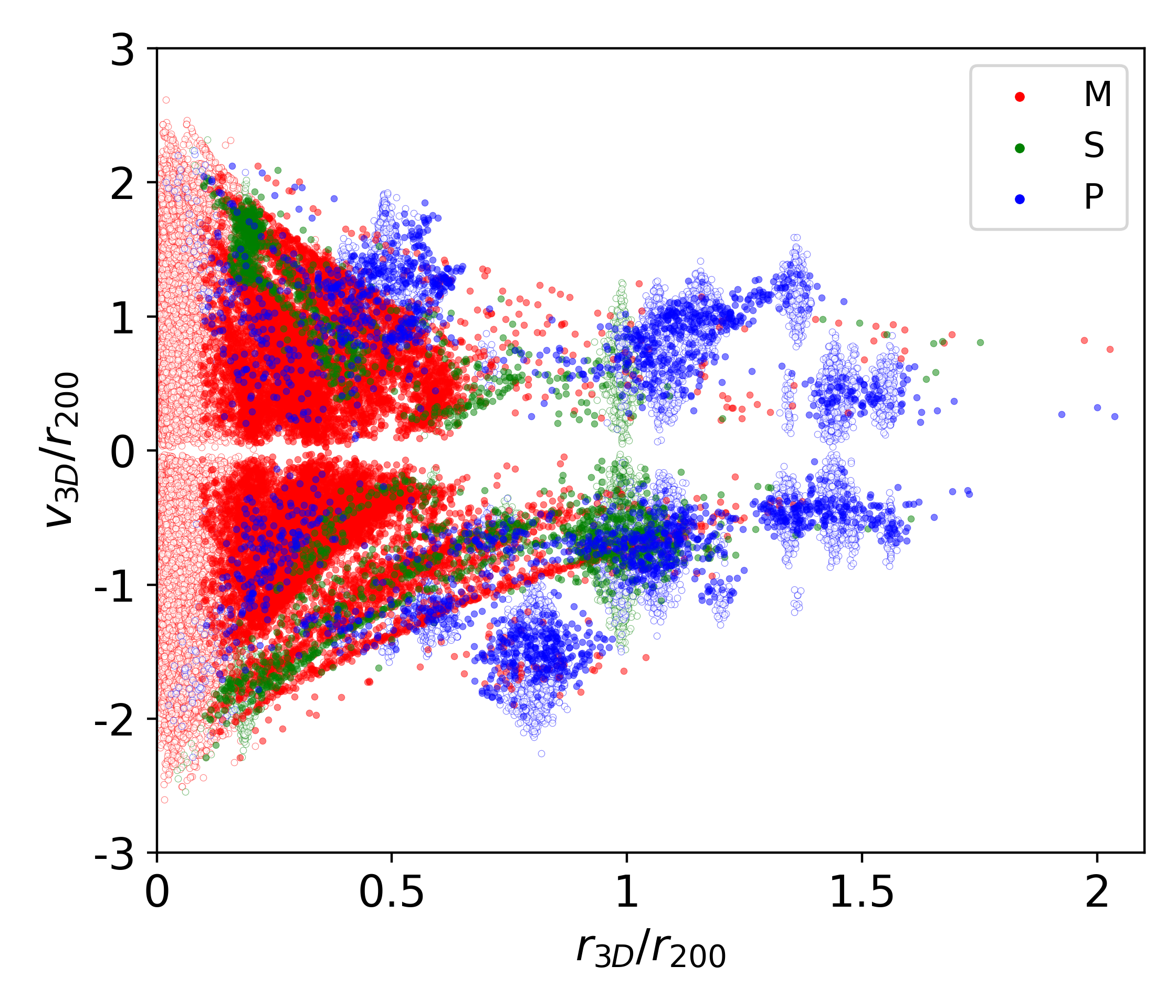}
    \caption{
    Phase-space diagram of all stellar particles from the halo displayed in Figure \ref{fig:galaxy_tree} at $z=0.625$.
    The open (filled) circles depict the bound (unbound) stellar particles.
    The red, blue, and green colors represent IHL stellar particles classified as \texttt{M}, \texttt{S}, and \texttt{P}, respectively.
    The \texttt{M} particles are cospatial with the BHG.
    The \texttt{S} particles form localized streaks connected to the BHG.
    The \texttt{P} particles are found around the bound particles of infalling galaxies.
    }
    \label{fig:phasespace}
\end{figure}

\subsubsection{Galaxy Chain Construction}
After constructing halo chains, we proceed to create galaxy chains within each halo in a similar way, except that we start with the highest redshift where the galaxy first appears within the halo.
Initially, the galaxy catalog for each halo is empty.
When a new galaxy is found, it is added to the galaxy catalog, and its chain is constructed as we visit lower and lower redshift snapshots.
We also collect the IDs of stellar particles bound to the galaxy and save the list.
We use the particle ID list to locate the child galaxy in the next (lower-redshift) snapshot.
We keep expanding the particle ID list of each galaxy by adding new particle IDs of the child galaxy as we build the chain.
In some cases, the stellar particles of a galaxy belong to multiple galaxies in the next snapshot, creating ambiguity.
In these cases, its child halo is determined as follows.
\begin{itemize}
    \item If all the candidate galaxies in the next snapshot have already been linked to their parent galaxies, we skip that snapshot.
    \item If there remains one galaxy that has not been linked to its parent galaxy, we choose that galaxy as the child galaxy.
    \item If more than one galaxy remains unlinked with their parent galaxies, we choose the most massive one as the child galaxy.
\end{itemize}
When we reach the last snapshot, the galaxy chain construction is completed. We then repeat the procedure for the next galaxy.
If we fail to find a child galaxy\footnote{This happens when it merges with another galaxy, escapes from the halo, and cannot be resolved by \texttt{PGalF}.}, we skip that snapshot and continue searching in the subsequent snapshots until reaching the final snapshot.
This effectively mitigates the impacts of any instability of \texttt{PGalF} on galaxy detection.

The top panel of Figure \ref{fig:galaxy_tree} displays an example of galaxy chains for one arbitrary halo.
The galaxy IDs are sorted according to the order of emergence within the halo.
In this example, galaxy 0, which seeds the BHG of the halo, formed first and maintains its chain to the lowest redshift.
However, many subsequently emerging galaxies discontinue and reappear for various reasons, including merging with another galaxy, escaping from the halo, and simple \texttt{PGalF} detection failure.

\subsection{Galaxy Status Classification}
\label{subsection:galaxy-classification}
To identify the origin of an IHL stellar particle, we need to determine its originating galaxy and ascertain its status at the time of identification.
We classify the galaxy status into the following three categories: merged with the BHG (\texttt{M}), stripped to the BHG (\texttt{S}), and infalling (\texttt{P}).
To facilitate the classification, we define $M_{i}(z)$, which is the mass of stellar particles bound to the i$^{th}$ galaxy at redshift $z$.
$M_{i}(z)$ excludes the masses of particles that do not originate from the i$^{th}$ galaxy, even if they are bound to the i$^{th}$ galaxy at $z$.
In other words, $M_{i}(z)$ includes only the masses of the stellar particles that have never been gravitationally bound to any other galaxies before.
Likewise, we define $M_{i}^{j}(z)$ as the mass of stellar particles that were initially bound to the $i^{th}$ galaxy and later transferred to the $j^{th}$ galaxy at redshift $z$.
The $i=j$ case represents the stellar mass remaining in the same galaxy.
Based on this migration history of stellar particles, we classify the $i^{th}$ galaxy status $S_i(z)$ at redshift $z$ into three categories as follows:
\begin{equation}
    S_{i}(z) =
    \begin{cases}
    \texttt{M}, &\text{if } M_{i}^{0}(z) \geq \alpha \sum_{j=0}^{n} M_{i}^{j}(z) \\
    \texttt{S}, &\text{else if } M_{i}^{0}(z) \geq \beta \sum_{j=1}^{n} M_{i}^{j}(z) (1-\delta_{ij}) \\
   \texttt{P}, &\text{otherwise},
    \end{cases}
    \label{equation:sh_class}
\end{equation}
where the superscript 0 refers to the BHG, and $n$ is the total number of galaxies within each halo.
In equation~\ref{equation:sh_class}, $\alpha$ is the threshold that determines the \texttt{M} status.  
We use $\alpha=0.5$ as a fiducial value for presenting the main results.
That is the galaxy status is considered as {\it merged with} the BHG, if more than 50\% of its stellar particle mass has contributed to the BHG.
However, in \textsection\ref{subsec:critera_uncertainty}, we demonstrate that our conclusion is not sensitive to the exact choice.
Likewise, $\beta$ determines the criterion for the \texttt{S} status. 
We adopt $\beta=1$. In other words, the $i^{th}$ galaxy status is considered as \texttt{S} if the status is not \texttt{M} and the summation of $M_i^j(z)$ over $j$, where we exclude the $j=0$ and $j=i$ cases, is less than the total stellar mass transferred to the BHG.
If the galaxy status is neither \texttt{M} nor \texttt{S}, we assign the \texttt{P} status to that galaxy.
The bottom panel of Figure \ref{fig:galaxy_tree} shows the status classification of the galaxies displayed in the top panel. 

Once the above galaxy status classification is completed, the IHL particle classification is straightforward.
For each IHL particle at $z$, we locate the originating galaxy, to which the particle was initially bound, and examine the galaxy status at the same redshift $z$.
If the galaxy statuses are \texttt{M}, \texttt{S}, and \texttt{P}, the IHL particle is considered to originate from the BHG formation, stripping, and pre-processing, respectively.
Figure \ref{fig:class_example} displays the spatial distribution of IHL particles originating from \texttt{M}, \texttt{S}, and \texttt{P} for the same four halos featured in Figure \ref{fig:halo_example}.
The spatial distributions of IHL particles with different origins are consistent with our expectations.
The IHL particles tagged \texttt{M} are mostly concentrated around the central region of each halo, where its BHG is located. 
In the case of the \texttt{S} class, stripping patterns are clearly visible.
The \texttt{P} particles tend to distribute around the member galaxies (except for BHGs) in both central and outskirt regions of halos.

The IHL particles with different origins are also clearly segregated in the phase-space diagram displayed in Figure \ref{fig:phasespace}.
The \texttt{M} particles are mostly concentrated in the inner part of the phase-space diagram without distinct clumpiness, while the \texttt{S} and \texttt{P} particles distribute in localized clumps across the radius.
The difference in distribution between the \texttt{S} and \texttt{P} classes is that the \texttt{S} particles tend to appear as narrow streaks associated with the BHG in the phase-space diagram, indicating their status having been stripped to the BHG.
The \texttt{P} particles are found around the bound particles of the member galaxies across the radius with no such features.
Figures~\ref{fig:class_example} and \ref{fig:phasespace} demonstrate that our IHL classification scheme, based on the status of the originating galaxy, effectively distinguish the IHL origins.

\begin{figure}
    \centering
    \includegraphics[width=0.95\columnwidth]{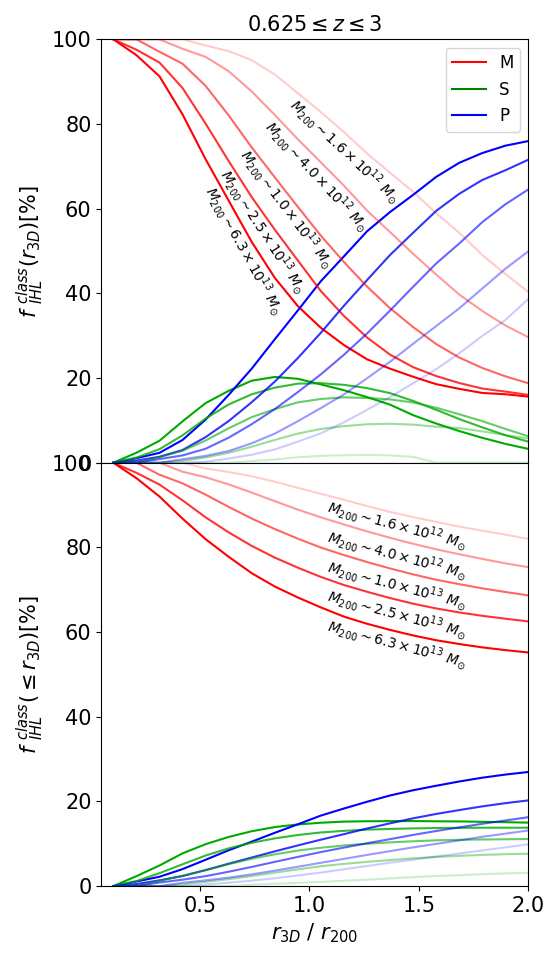}
    \caption{
    Radial profiles of $f_{\mathrm{IHL}}^{class}$.
    We use red, green, and blue for the \texttt{M}, \texttt{S}, and \texttt{P} channels, respectively.
    Within each channel, the intensity of the curves denotes different mass bins, with higher intensities corresponding to more massive halos.
    The median of each mass bin is displayed below the corresponding curve.
    (Top) 
    $f_{\mathrm{IHL}}^{class} (r)$ versus radius.
    The \texttt{M} channel is the dominant IHL production mechanism at small radii, while the  \texttt{P} channel becomes more important at large radii.
    The contribution from the \texttt{S} channel peaks at intermediate radii, although its maximum remains below $\sim 20 \%$.
    (Bottom) Cumulative profile.
    The \texttt{M} channel is the primary contributor to the total IHL at all radii.
    The fractions of both the \texttt{S} and \texttt{P} channels increase with radius, but $f_{\mathrm{IHL}}^{\texttt{S}}$ levels off at larger radii while $f_{\mathrm{IHL}}^{\texttt{P}}$ continues to increase.
    While the contributions from the \texttt{P} and \texttt{S} channels increase with halo mass, the \texttt{M} channel exhibits the opposite behavior.
    }
    \label{fig:radius_distribution}
\end{figure}

\section{Result}
\label{sec:result}

\subsection{Radial Distribution Analysis}
\label{subsection:distribution_of_IHL}

\begin{figure*}
    \centering
    \includegraphics[width=0.49\textwidth]{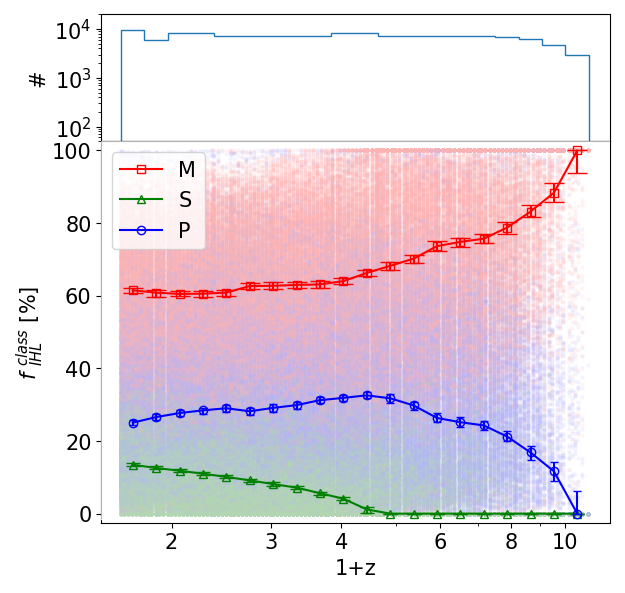}
    \hfill
    \includegraphics[width=0.49\textwidth]{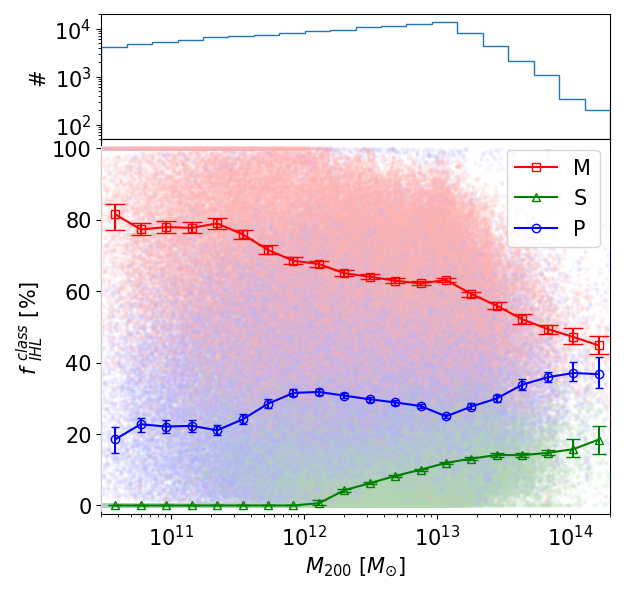}
    \caption{
    Evolution of $f^{class}_\mathrm{IHL}$.
    The histograms show the number of halos per each bin.
    Each dot represents $f^{class}_\mathrm{IHL}$ of each halo.
    The red squares, green triangles, and blue circles represent the medians of $f_\mathrm{IHL}^{M}$, $f_\mathrm{IHL}^{S}$, and $f_\mathrm{IHL}^{P}$, respectively, within each bin.
    The error bar depicts three times the standard error of the median estimated with the bootstrap resampling method.
    (Left) $f^{class}_\mathrm{IHL}$ with redshift.
    $f_\mathrm{IHL}^{M}$ peaks at the highest redshift, decreases with age, and levels off at $z\lesssim3$ at $\mytilde60$\%.
    $f_\mathrm{IHL}^{P}$ starts to increase rapidly from $\mytilde0$\% at $z=9$, reaches its maximum $\mytilde30$\% at $z=3\sim4$, and slowly decreases with age to $\mytilde25$\% at the last snapshot $z=0.625$.
    $f_\mathrm{IHL}^{S}$ remains negligible until $z\gtrsim4$ and steadily increases with age to $\mytilde15$\% at $z=0.625$.
    (Right)  $f^{class}_\mathrm{IHL}$ as a function of halo mass.
    $f_\mathrm{IHL}^{M}$ is $\mytilde 80\%$ at $M_{200} \mytilde 10^{11} M_{\odot}$ and decreases to $\mytilde 45\%$ at $M_{200} \sim 10^{14} M_{\odot}$, while in the same mass range $f_\mathrm{IHL}^{P}$ increases from $\mytilde20\%$ to $\mytilde40\%$.
    $f_\mathrm{IHL}^{S}$ emerged at $M_{200} \gtrsim 10^{12} M_{\odot}$ and reached to $\mytilde15\%$ at $M_{200} \sim 10^{14} M_{\odot}$.
    One may interpret the trends around $M_{200} \sim 10^{14} M_{\odot}$ as indicating a crossover of the dominant IHL formation channel from \texttt{M} to \texttt{P}.
    However, caution is warranted due to limited statistics in this mass regime. (See text)}
    \label{fig:zorigins}
\end{figure*}

In \textsection\ref{subsection:galaxy-classification}, we briefly discussed how differently the three classes (\texttt{M}, \texttt{S}, and \texttt{P}) of IHL distribute qualitatively for a few examples.
In this section, we present the radial distribution estimated with a larger HR5 sample.
We define the IHL class fraction $f^{class}_{\mathrm{IHL}}$ as follows:
\begin{equation}
    {f_{\mathrm{IHL}}^{class}} (r) = \frac{M_{\mathrm{IHL}}^{class}(r)}{M_{\mathrm{IHL}}(r)},
\end{equation}
where the \texttt{class} superscript is \texttt{M}, \texttt{S}, or \texttt{P}, and $M_{\mathrm{IHL}}(r)$ is the total mass of all IHL stellar particles at $r$.
$M_{\mathrm{IHL}}^{class}(r)$ is the same as $M_{\mathrm{IHL}}(r)$ except that it includes only the IHL particles with a specific class.

Figure \ref{fig:radius_distribution} shows the radial distribution of $f^{class}_{\mathrm{IHL}}$. We limit the redshift range of the sample to $0.625\leq z \leq 3$.
As we showed in the top panel, the \texttt{M} type IHL particles dominate the central region ($r \lesssim r_{200}$) across all mass bins, their contribution steadily decreases with increasing radius.
Its dominance in the central region is more pronounced in lower-mass halos.
The \texttt{P} class shows a continuous rise with radius and becomes a dominant component in the outskirts ($r \gtrsim 1.5\ r_{200}$), especially in higher mass.
The \texttt{S} class exhibits a moderate increase toward $r \sim r_{200}$, beyond which it gradually decreases.
The higher-mass halos show earlier and more pronounced turnover with higher amplitudes.
The cumulative profiles show that the \texttt{M} class remains as the most dominant contributor to the total IHL budget at all radii, despite the trend observed in the top panel, where contributions from the other channels surpass $f_{\mathrm{IHL}}^{M}(r)$ in the outer regions.

The shapes of these radial profiles reflect the distinct physical processes governing each channel.
The finding that IHL formation via mergers with BHGs peaks in the central regions is expected, as this process is most active there.
At larger radii, IHL production on group scales is more prominent, reflecting the increased presence of infalling groups.
As we treat stripping and mergers as processes along a continuum rather than entirely distinct mechanisms in the current study, IHL production via stripping is most pronounced during the phase when satellite galaxies orbit the BHG before eventually merging, resulting in a strongest contribution at intermediate radii.
More massive halos tend to host more satellites and infalling groups \citep{Berlind2002}, thereby enhancing both stripping and pre-processing contributions.

\subsection{\texorpdfstring{Evolution of $f^{class}_{\mathrm{IHL}}$}{Evolution of f class IHL with Redshift}}
\label{subsection:IHL_evolution}

In \textsection\ref{subsection:distribution_of_IHL}, we show that among the three IHL production channels, the merger with the BHG is the most important for the cluster as a whole, followed by the pre-processing.
In this section, we examine how this trend varies quantitatively with redshift and halo mass. 
The left panel of Figure~\ref{fig:zorigins} displays the evolution of $f_{\mathrm{IHL}}^{class}$ with redshift.
At $z\sim10$, the dominance of the \texttt{M} channel is extreme.
As the universe evolves, $f_{\mathrm{IHL}}^{M}$ gradually decreases and nearly levels off at $z\lesssim3$.
The contribution from the \texttt{P} channel rapidly rises from $z\sim10$ and reaches $f_{\mathrm{IHL}}^{P}\sim30$\% at $z=3\sim4$.
At $z\lesssim3$, it slowly decreases, reaching $\mytilde25$\% at $z=0.625$.
The \texttt{S} channel contribution stays negligible from $z\sim10$ to $z\sim4$, from which it gradually increases and becomes $f_{\mathrm{IHL}}^{S}\sim15$\% at $z=0.625$.

The trends described above show that the IHL production channel through the formation of BHGs is very active in the early ($z\gtrsim4$) universe.
We propose that this is because the merger rate was very high in the early, dense universe.
The rapid increase in the contribution from the \texttt{P} channel indicates that the \texttt{P} channel might be  similar to the \texttt{M} channel, except that the mergers happen to massive galaxies that are infalling into more massive halos.
Stripping becomes important only after clusters become sufficiently massive at $z\lesssim3$, when their tidal forces become significant.
Since HR5 stops at $z=0.625$, it is not clear whether the \texttt{S} channel production rate would overtake the \texttt{P} channel production rate at $z\sim0$.
Nevertheless, rough extrapolation suggests that it is very likely that the \texttt{M} channel remains dominant down to $z=0$.

The right panel of Figure \ref{fig:zorigins} shows that $f_{\mathrm{IHL}}^{class}$ also significantly varies with halo mass. 
$f_{\mathrm{IHL}}^{M}$ is $\mytilde 80\%$ at $M_{200} \mytilde 10^{11} M_{\odot}$ and decreases with halo mass, reaching $\mytilde 40\%$ at $\mytilde 10^{14} M_{\odot}$. 
In the same mass range, $f_{\mathrm{IHL}}^{P}$ increases from $\mytilde20\%$ to $\mytilde30\%$.
$f_{\mathrm{IHL}}^{S}$ remains negligible below $M_{200} \lesssim 10^{12} M_{\odot}$ and increases up to $~10\%$. 
As can be seen in the histogram, the number of halos precipitously decreases at $M_{200} \gtrsim 10^{13} M_{\odot}$, which calls for caution when interpreting the results in this mass regime.
The statistics at $M_{200} \gtrsim 10^{13} M_{\odot}$ are dominated by samples from the last few snapshots in HR5.
Since the HR5 simulation stops at $z=0.625$, the results are biased due to the absence of halo contributions at $z<0.625$.
If HR5 were complete down to $z=0$, the decline of $f_{\mathrm{IHL}}^{M}$ at $M_{200} \gtrsim 10^{13} M_{\odot}$ might have been more gradual, resulting in a larger gap between $f_{\mathrm{IHL}}^{M}$ and $f_{\mathrm{IHL}}^{P}$ around $M_{200} \sim 10^{14} M_{\odot}$.
We discuss this issue further in Section \ref{subsec:highmassbias}.

\section{Discussion}
\label{sec:discussion}

\subsection{Sample bias in mass dependence}
\label{subsec:highmassbias}

\begin{figure}
    \centering
    \includegraphics[width=0.98\textwidth]{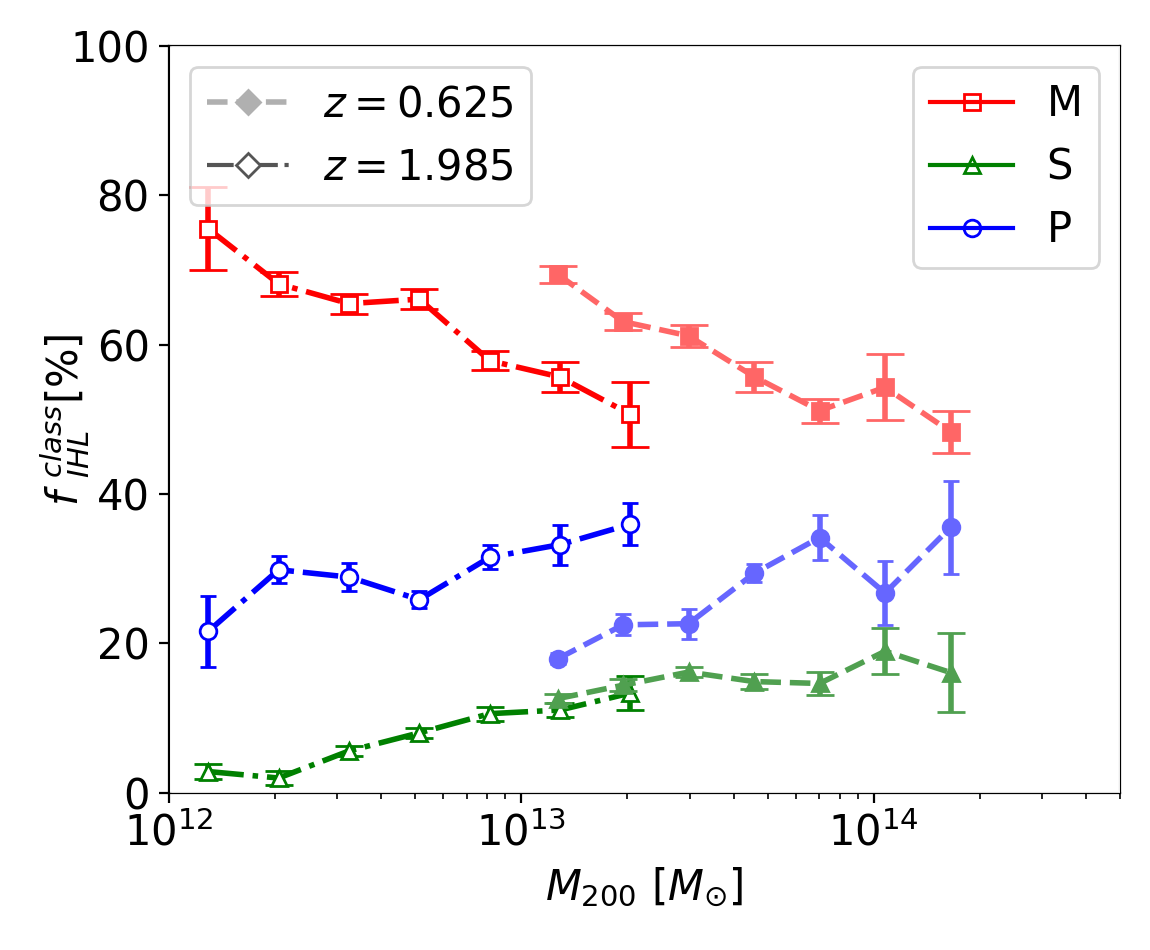}
    \caption{
    $f_{\mathrm{IHL}}^{\mathrm{class}}$ as a function of halo mass at $z=0.625$ and $z=1.985$.
    The same color and symbol schemes as in Figure \ref{fig:zorigins} are used, with filled (open) symbols representing $z=0.625$ ($z=1.985$).
    Both snapshots show that $f_{\mathrm{IHL}}^{M}$ is higher in lower mass halos, while the trend is opposite for $f_{\mathrm{IHL}}^{P}$.
    However, at the same halo mass (e.g. $\mytilde 2\times10^{13} M_{\odot}$), a systematic offset is present between the two snapshots.
    On the other hand, $f_{\mathrm{IHL}}^{S}$ increases gradually with halo mass and no systematic difference between two snapshots exists.
    }
    \label{fig:revisit}
\end{figure}

As discussed in Section \ref{subsection:IHL_evolution}, the apparent mass dependence in the $M_{200} \gtrsim 10^{13} M_{\odot}$ regime may be subject to bias arising from the finite redshift extent of the HR5 dataset.
Since the simulation ends at $z = 0.625$, massive halos at $z<0.625$ are not included.
To illustrate this effect, we compare the mass dependence of $f_{\mathrm{IHL}}^{class}$ at $z = 0.625$ and $z = 1.985$ in Figure \ref{fig:revisit}.
We note that, as this is a longitudinal study tracing the same halo over time, the two redshift samples largely comprise the same halos, and the differences primarily reflect genuine time evolution.

At both $z = 1.985$ and $z = 0.625$, the \texttt{M} and \texttt{P} channels show a clear dependence on halo mass.
$f_{\mathrm{IHL}}^{M}$ at $z=1.985$ decreases from $\mytilde75\%$ at $M_{200} \sim 10^{12} M_\odot$ to $\mytilde50\%$ at $2 \times 10^{13} M_\odot$, while $f_{\mathrm{IHL}}^{P}$ increases from $\mytilde20\%$ to $\mytilde35\%$ in the same range. 
At $z = 0.625$, we found a similar trend, with $f_{\mathrm{IHL}}^{M}$ declining from $\mytilde70\%$ to $\mytilde45\%$ and $f_{\mathrm{IHL}}^{P}$ rising from $\mytilde20\%$ to $\mytilde35\%$, across a higher mass range of $10^{13} M_\odot \lesssim M_{200} \lesssim2 \times 10^{14} M_\odot$.
Since the variations in $f_{\mathrm{IHL}}^{class}$ at the two redshifts span similar ranges across different mass intervals, significant differences emerge within the overlapping mass range ($M_{200} \sim 2 \times 10^{13} M_{\odot}$). 
Had HR5 terminated at $z=1.985$, $f_{\mathrm{IHL}}^{M}$ ($f_{\mathrm{IHL}}^{P}$) in the right panel of Figure \ref{fig:zorigins} would have been biased low (high) around $\mytilde 10^{13} M_{\odot}$, due to the absence of contributions from $z<1.985$.
The same reasoning applies to the present case, where the final snapshot of HR5 is at $z=0.625$.
Therefore, caution is warranted when interpreting the results at $M_{200} \gtrsim 10^{13} M_{\odot}$, where similar bias is expected to be present.

In contrast, the \texttt{S} channel shows a negligible difference between the two snapshots at $M_{200} \sim 2 \times 10^{13} M_\odot$.
This suggests that the mass dependence observed for the \texttt{S} channel may be less susceptible to bias arising from limited snapshot coverage.
Further analysis is required to understand why the variation of $f_{\mathrm{IHL}}^{S}$ remains continuous across the mass range at different redshifts.

\subsection{Impacts of Classification Criteria}
\label{subsec:critera_uncertainty}
\begin{figure}
    \centering
    \includegraphics[width=\columnwidth]{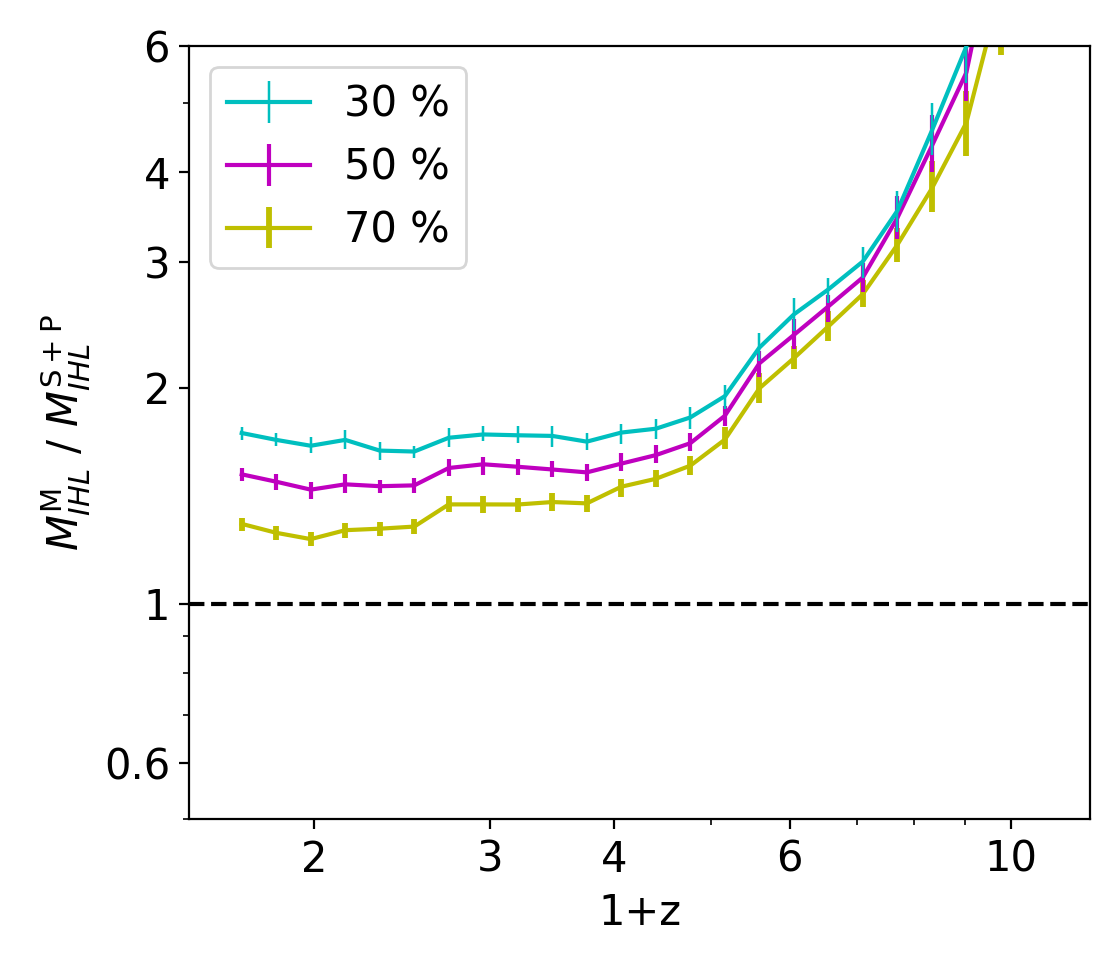}
    \caption{Impact of the classification criteria.
    We compared the mass of the IHL formed through the \texttt{M} channel $M^{M}_{\mathrm{IHL}}$ with the IHL mass produced by the sum of the other two classes $M^{S+P}_{\mathrm{IHL}}$ by changing the coefficient $\alpha$ in Equation \ref{equation:sh_class}.
    The cyan, magenta, and yellow lines show the $M^{M}_{\mathrm{IHL}} / M^{S+P}_{\mathrm{IHL}}$ values when the coefficients are set to 0.3, 0.5, and 0.7, respectively.
    Setting it at 0.3 increases the fraction $M^{M}_{\mathrm{IHL}}/M^{S+P}_{\mathrm{IHL}}$   by $\mytilde12\%$ while setting it at 0.7 decreases the fraction by $\mytilde12\%$.
    However, it is important to note that the overall trend and interpretation of the results remain consistent regardless of the exact value chosen for the coefficient.
    }
    \label{fig:uncertainty_criteria}
\end{figure}

From a broad perspective, merging and stripping can be considered physically identical processes viewed at different epochs.
In the early stage when a stripped galaxy still maintains a significant fraction of its mass, the resulting IHL stars are regarded as originating from stripping.
However, in the late stage, when the same galaxy has lost most of its mass, all the IHL stars produced up to that point should be attributed to a merger.

In equation~\ref{equation:sh_class}, we adopt $\alpha=0.5$ for the fiducial case, implying that the critical mass fraction distinguishing merging from stripping is 50\%.
Of course, this choice is arbitrary.
One may argue that the result could vary significantly if a different value were chosen.
In this section, we explore how the result changes for a different choice of $\alpha$.

Figure \ref{fig:uncertainty_criteria} displays the evolution of the ratio $M^M_{\mathrm{IHL}}/M^{S+P}_{\mathrm{IHL}}$ across redshift for $\alpha=0.3$, $0.5$, and $0.7$.
As expected, the $M^M_{\mathrm{IHL}}/M^{S+P}_{\mathrm{IHL}}$ ratio increases (decreases) when  $\alpha=0.3$ (0.7), as a lower (higher) value regards the IHL production as originating from merging at an earlier (later) stage.
However, our conclusion that merging is still the most dominant IHL production channel remains unchanged.
At the last HR5 snapshot ($z=0.625$), the $M^M_{\mathrm{IHL}}/M^{S+P}_{\mathrm{IHL}}$ ratio approximately doubles when $\alpha$ changes from 0.3 to 0.7. 
Toward higher redshifts, the difference decreases.

\begin{figure}
    \centering
    \includegraphics[width=\columnwidth]{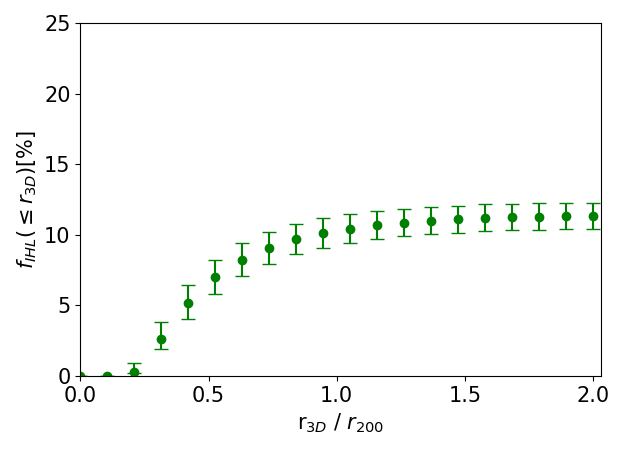}
    \caption{
    Dependence of the cumulative IHL fraction $f_{\mathrm{IHL}}(r\leq r_{3D})$ on the reference radius.
    Green dots denote the median cumulative IHL fraction at $r<r_{3D}/r_{200}$ within each radial bin. 
    The error bars indicate 100 times the standard error of the median.
    The IHL fraction is zero at small radii, where all the stars are bound to the BHG.
    The fraction rapidly goes up until $r\sim r_{200}$, where it starts to level off at $\mytilde10$\%.
    The mean IHL fraction is sensitive to the choice of the radius cut if $r\lesssim r_{200}$.
    }
    \label{fig:rfractions}
\end{figure}

\subsection{Radial Dependence of IHL Fraction}
\label{subsection:IHLfraction_inandout}
A difficulty in comparing different IHL studies is the choice of radius within which the IHL is quantified. 
Here we investigate the IHL fraction as a function of radius normalized by $r_{200}$.
We define a cumulative IHL fraction $f_{\mathrm{IHL}} (\leq r)$ as follows:
\begin{equation}
    f_{\mathrm{IHL}} (\leq r) = \frac{M_{\mathrm{IHL}} (\leq r)}{M_{\mathrm{star}} (\leq r)},
    \label{equation:frac_in}
\end{equation}
where $r$ is the cluster centric radius and $M_{\mathrm{star}} (\leq r)$ is the total stellar mass within $r$.
Figure \ref{fig:rfractions} shows the mean cumulative IHL fraction as a function of the normalized radius.
The IHL fraction is zero at small radii, where there exist only the stars bound to the BHG.
The fraction rapidly rises until $r\sim r_{200}$, where it starts to level off at $\mytilde10$\%.
This demonstrates that the mean IHL fraction is sensitive to the choice of the radius cut if $r\lesssim r_{200}$. 

\subsection{IHL Contribution from Dwarf Disruption and In-situ Star formation}
\begin{figure}
    \centering
    \includegraphics[width=\columnwidth]{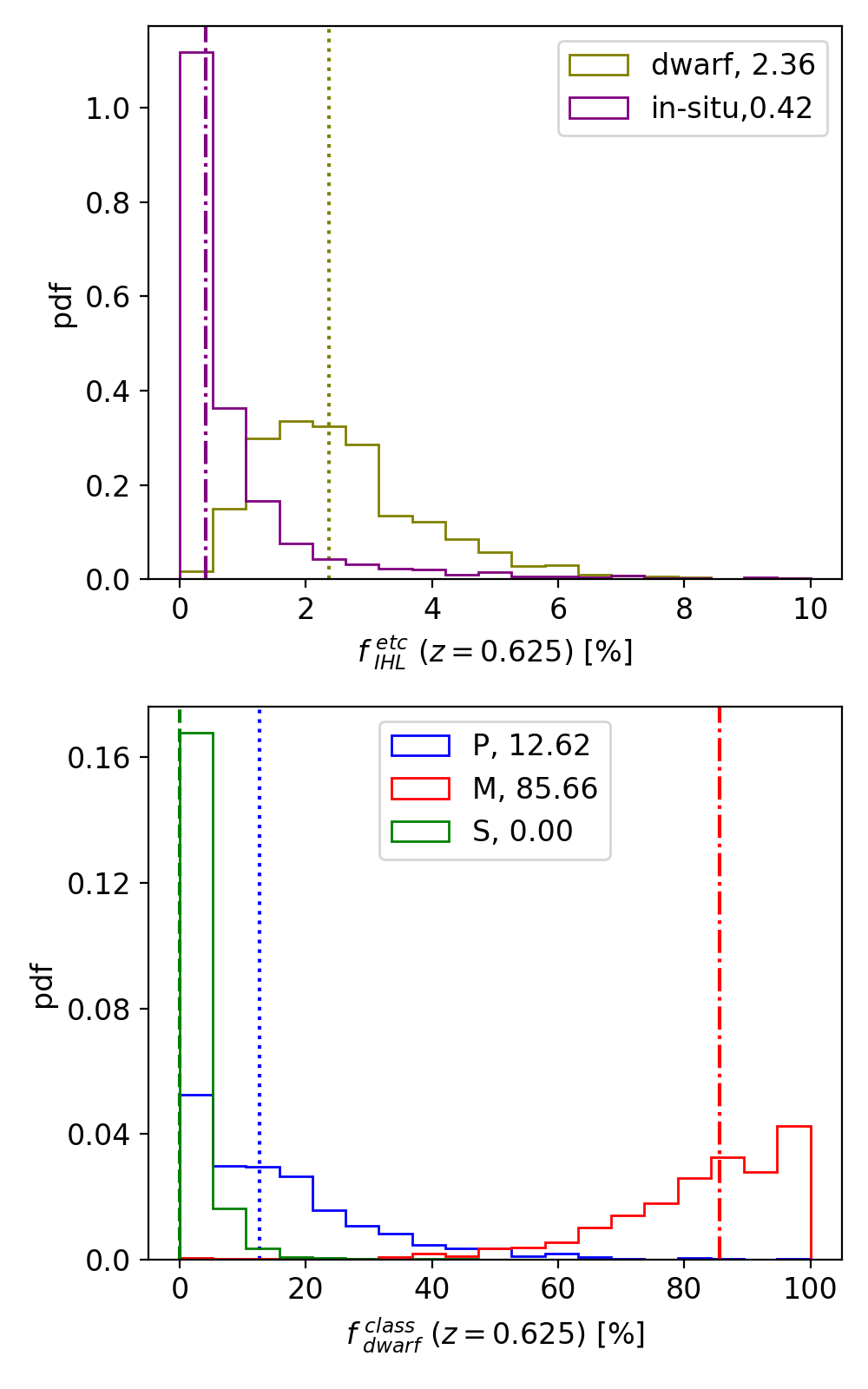}
    \caption{
    Fraction of dwarf disruption or in-situ formation at $z=0.625$ based on 1,202 halos.
    (Top) The distribution of the  mass fraction ($f_{IHL}$) of IHL originating from dwarf galaxies (olive green) or in-situ star formation (purple) is illustrated.
    The median of $f_{IHL}$ from dwarf disruption (in-situ star formation) is 2.36\% (0.37\%) of the total IHL.
    Both mechanisms contribute minimally to the total IHL production.
    (Bottom) The distribution of the fraction of IHL originating from dwarf galaxies but (incorrectly) classified into the three main
    (\texttt{M}, \texttt{P}, and \texttt{S}) channels is illustrated.
    This indicates that the dwarf channel induces $\mytilde2\%$ positive bias in the \texttt{M}-type channel.
    }
    \label{fig:dwarf_and_insitu}
\end{figure}

Until now, we have mainly focused on the three primary mechanisms for the production of IHL: BHG formation, stripping, and pre-processing.
In this section, we examine two additional mechanisms.

The first one is the disruption of dwarf galaxies. Dwarf galaxies are vulnerable to tidal forces due to their low mass and shallow gravitational potential wells.
When these dwarf galaxies pass close to the cluster center or interact with larger galaxies, the tidal forces can completely disrupt them.
This disruption results in their stars becoming unbound, subsequently contributing to the IHL.
To trace the IHL originating from dwarf galaxies, we investigate galaxies at $z=0.625$ with stellar masses of $10^{9} \mathrm{M_{\odot}}$ or less.
In the top panel of Figure \ref{fig:dwarf_and_insitu}, the distribution of
$f_{IHL}$, the mass fraction of IHL originating from dwarf galaxies with respect to the total IHL mass, is shown in olive green.
The $f_{IHL}$ distribution shows that the IHL originating from dwarf galaxies is very small (the median fraction is 2.36\%).
This result aligns well with previous studies \citep[e.g.,][]{Contini2014, DeMaio2015}.

We extended this study by quantifying the fraction of IHL originating from dwarf galaxies but (incorrectly) classified into the three main channels
(\texttt{M}, \texttt{S}, or \texttt{P}).
The bottom panel of Figure \ref{fig:dwarf_and_insitu} shows that over 85 percent of IHL produced from dwarf galaxies is classified as \texttt{M}, while approximately 13 percent is classified as \texttt{P}. The classification into \texttt{S} is negligible.
This illustrates that about 2\% positive bias is introduced in our classification of the \texttt{M}-channel contribution.

The second mechanism is in-situ star formation, referring to the production of stars within the intrahalo medium without being bound to any particular galaxy.
The methods for tracing in-situ stars vary across studies.
For instance, \cite{Brown2024} considered stars that had never been part of any galaxies.
On the other hand, \cite{ahvazi2024} included  stars that had formed within the BHG but were later pushed onto highly energetic orbits due to mergers.
In this study, we trace stars that had never been bound to any galaxies.

The purple histogram in the top panel of Figure \ref{fig:dwarf_and_insitu} illustrates the fraction of in-situ IHL mass relative to the total IHL mass.
We find that less than 1\% of the IHL was formed through the in-situ process.
This is well aligned with the findings of \cite{Brown2024} and \cite{Melnick2012}.

In summary, the disruption of dwarf galaxies contributes minimally, accounting for a median of 2.36\% of the total IHL, while in-situ star formation contributes even less, accounting for less than 1\%.
These findings confirm that these two processes are minor contributors to the total IHL formation compared to the other three main mechanisms.

\section{Conclusion}
\label{sec:concl}

We have investigated the formation and evolution of IHL using the HR5 cosmological simulation, which offers the large volume and high resolution essential for reliable identification and statistically robust analysis of unbound star particles.
IHL star particles were identified utilizing the \texttt{PGalF} algorithm, which employs both binding energies and tidal radii to differentiate unbound and bound stellar particles of galaxies.
Halo chains and galaxy chains were constructed by tracing their member particles and positions. 

We keep track of individual particles' gravitational binding and transfer histories and use this information to classify each galaxy's status at each redshift into the three categories: BHG merging, stripping, and pre-processing. These three IHL production channels are not regarded as distinct processes but as points on a continuum, depending on the degree of stellar particle transfer.
This provides the foundation that enables us to reliably and quantitatively identify the production channel for each IHL star particle.
Our main findings are summarized as follows.

{\bf Main production channel:}
\begin{itemize}
\item Since large halo-to-halo variations in IHL production mechanisms exist, caution should be used when interpreting results from a small sample.
\item On average, the formation and growth of BHGs are the predominant mechanism for IHL production, accounting for the largest fraction of IHL mass.
\item The IHL stars produced through this BHG formation mechanism are mostly concentrated around BHGs, with occasional extensions to the outskirts of the halos.
\item As halos evolve, the fraction of IHL originating from BHG mergers decreases slightly but remains the most significant contributor, with $\geq 60\%$ even in the last HR5 snapshot ($z=0.625$).
\item Pre-processing, the IHL production prior to infall into the main halo on a group scale, also contributes significantly ($\mytilde20$\%) to the total IHL production budget.
\item Pre-processing becomes more important as the number of infalling galaxies increases with time.
\item IHL production through stripping also plays a non-negligible role, particularly at low redshifts, as tidal stripping of satellite galaxies by the halos' potential becomes efficient.
\item The relative contributions of the three channels vary with halo mass, with \texttt{M} channel becoming less dominant at higher masses, while the \texttt{P} and \texttt{S} channels gain in importance.
Nevertheless, the hierarchical order ($f_{\mathrm{IHL}}^{M} > f_{\mathrm{IHL}}^{P} > f_{\mathrm{IHL}}^{S}$) remains valid across the explored mass range.

\end{itemize}

{\bf Radial properties:}
\begin{itemize}
\item The IHL originating from BCG mergers is highly concentrated in the central regions of the halos.
\item IHL produced via satellite stripping and pre-processing becomes more prominent at larger radii, indicating ongoing dynamical interactions and hierarchical build-up of galaxy clusters.
\item The fraction of IHL resulting from pre-processing continues to increase with radius, suggesting that these IHL stars are indeed associated with infalling galaxies.
\item The IHL fraction increases with radius until it levels off at $\mytilde11$\% at $r_{3D} \geq r_{200}$, suggesting that the amount of IHL production through pre-processing on a group/protocluster scale is substantial.
\end{itemize}

{\bf Dwarf disruption and In-situ formation:}
\begin{itemize}
\item The disruption of dwarf galaxies contributes a median of 2.36\% to the IHL mass at $z=0.625$, aligning with past studies, and most of IHL is classified as \texttt{M} type in our analysis.
\item Less than 1\% of IHL forms directly in intra-halo medium, with variation in methods across studies.
\item Both mechanisms contribute minimally to IHL formation compared to the three primary mechanisms focused on in this study.
\end{itemize}

{\bf IHL fraction evolution:}
\begin{itemize}
\item The evolution of the IHL fraction is nearly flat at $f_{\mathrm{IHL}}\sim0.12$ in the $z\lesssim4$ regime, consistent with the recent observational results.
\item The analysis in the $4\lesssim z \lesssim 8$ regime suggests a potential increase in the IHL fraction with redshift.
\end{itemize}

Our study presents comprehensive theoretical insights into the major IHL production channels.
It also offers a few testable predictions, including the evolution of IHL fraction, that can be compared against current and future observations.
Data from facilities such as JWST and the Rubin observatory will be particularly valuable for disentangling the various contributions to IHL across different scales.
In particular, multi-band and spectroscopic observations will help constrain the properties of the underlying stellar populations, which are closely linked to the origins of the IHL.
We therefore anticipate that this framework will serve as a valuable reference for interpreting future observational analyses.

M. J. Jee acknowledges support for the current research from the National Research Foundation (NRF)
of Korea under the programs 2022R1A2C1003130 and RS-2023-00219959. This work benefited from the outstanding support provided by the KISTI National Supercomputing Center and its Nurion Supercomputer through the Grand Challenge Program (KSC- 2018-CHA-0003, KSC-2019-CHA-0002). Large data transfers were supported by KREONET, which is managed and operated by KISTI. This work is supported by the Center for Advanced Computation at Korea Institute for Advanced Study.
J. Lee is supported by the National Research Foundation of Korea (NRF-2021R1C1C2011626).
Y. Kim is supported by Korea Institute of Science and Technology Information (KISTI) under the institutional R\&D project (K25L2M2C3).

\appendix
\section{Comparison of IHL Fraction From Mass and Photometric Measurements}
\label{appendix:ihl_fraction_comparison}

\begin{figure}
    \centering
    \includegraphics[width=1.0\linewidth]{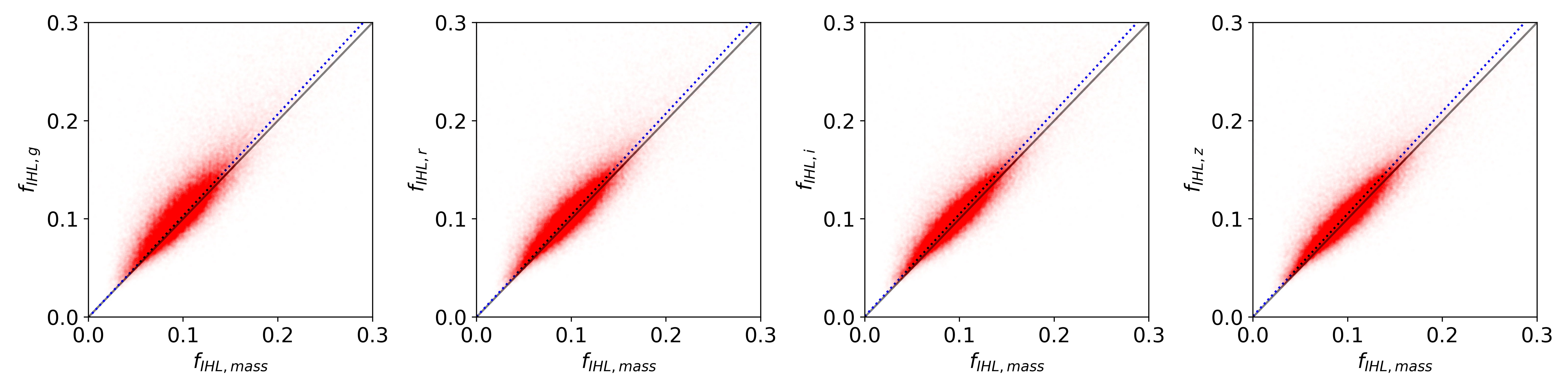}
    \caption{
    Comparison between the IHL fractions derived from stellar mass and those derived from synthetic photometry in the $g$, $r$, $i$, and $z$ bands. 
    Each red point represents a halo, the solid line indicates the one-to-one relationship, and dotted line shows fitted relation of each. 
    The slope of the fitted relation is 1.035, 1.034, 1.040, and 1.042 in the $g$, $r$, $i$, and $z$ bands, respectively, with all values lying within 5\% of unity.
    }
    \label{fig:scatter_figure}
\end{figure}

To test whether the mass-based IHL fraction (Section \ref{subsec:fraction_evolution} and Figure \ref{fig:fraction_IHL_comparison}) yields results consistent with flux-based estimates, we perform synthetic photometry in the  $g$, $r$, $i$, and $z$ bands and compare the resulting flux-based measurements with the mass-based ones.
These estimates are obtained using the SKIRT radiative transfer code \citep{Camps2020}, adopting the SDSS filter transmission curves.
Figure \ref{fig:scatter_figure} shows the comparison between the $f_{\mathrm{IHL}}$ values estimated from stellar mass and flux.
The photometry-based estimates in all four bands exhibit consistency with the mass-based values, with best-fit slopes of 1.035, 1.034, 1.040, and 1.042 in the $g$, $r$, $i$, and $z$ bands, respectively.

\bibliography{references}

\bibliographystyle{aasjournal}

\end{document}